\documentclass[printer]{aa}
\usepackage{color}
\usepackage{soul}
\setulcolor{red}
\usepackage{txfonts}
\usepackage{natbib}
\usepackage{graphicx}

\begin{document}

\title{Planetary nebula kinematics in NGC 1316: a young Sombrero \thanks{Based on observations made with the VLT at Paranal Observatory under program IDs 070.B-0089A}}
\author{E. K. McNeil-Moylan\inst{1,2,3}  \and K.C. Freeman\inst{1} \and M. Arnaboldi\inst{2} \and O.E. Gerhard\inst{3}}

\institute{Mt Stromlo Observatory, Australian National University, Cotter Road, Weston Creek, ACT 2611, Australia;
E-mail: emcneil@mso.anu.edu.au \and European Southern Observatory, Karl-Schwarzschild-Str. 2, D-85748 Garching, Germany; \and Max Planck Institute for Extraterrestrial Physics,
Karl-Schwarzschild-Str. 1, 85741 Garching, Germany;}

\date{Received date:  / 
Accepted date: }

\abstract    % context heading (optional)
	{}
  % aims heading (mandatory)
   {We present positions and velocities for 796 planetary nebulae (PNe) in the Fornax Brightest Cluster Galaxy NGC 1316 (Fornax A). The planetary nebulae and existing kinematics are used to explore the rotation of this merger remnant and constrain dynamical models.}
  % methods heading (mandatory)
   {Using FORS2 on the VLT, the PN velocities were measured using a counter-dispersed slitless-spectroscopy technique that produced the largest-to-date sample outside of the Local Group.  Spherical, non-rotating, constant-anisotropy Jeans models were constrained by observations of the planetary nebulae and  existing integrated light spectra.}
  % results heading (mandatory)
   {The two-dimensional velocity field indicates dynamically-important rotation that rises in the outer parts, possibly due to the outward transfer of angular momentum during the merger. The modeling indicates a high dark matter content, particularly in the outer parts, that is consistent with previous estimates from dynamical models, lensing and stellar population models. }
  % conclusions heading (optional), leave it empty if necessary  
	{The exceptionally large sample of PN velocities makes it possible to explore the kinematics of NGC 1316 in detail. Comparing the results to other early-type galaxies like NGC 1399 and NGC 4594 (M104, Sombrero), NGC 1316 represents a transition phase from a major-merger event to a bulge-dominated galaxy like NGC 4594. }


\titlerunning{Planetary Nebula Kinematics in NGC 1316}
\authorrunning{McNeil-Moylan \it{et al.}}

\maketitle
%%%%%%%%%%%%%%%%%%%%%%%%%%%%%%%%%%%%%%%%%%%%%%%%
%INTRODUCTION
%%%%%%%%%%%%%%%%%%%%%%%%%%%%%%%%%%%%%%%%%%%%%%%%
\section{Introduction}
Early-type galaxies hold an important place in the evolution of structure in the Universe, but their formation and development are not well understood. The halos are of particular interest because of the high dark-matter content and the link between the orbital structure and the formation history. However, these regions are difficult to observe because integrated light spectra fail to reach sufficient signal-to-noise at large radii. This challenge has been addressed using novel techniques on increasingly large telescopes \citep{Proctor:2009}, IFU spectroscopy \citep{Weijmans:2009}, and discrete tracers like globular clusters \citep{Spitler:2006, Schuberth:2010} and planetary nebulae. 

Planetary nebulae (PNe) provide a powerful tool for exploring elliptical stellar systems beyond 2-3R$_e$ \citep{Hui:1995, Arnaboldi:1998, Peng:2004, Mendez:2001, Mendez:2008, Doherty:2009, Coccato:2009, McNeil:2010}. They are bright enough to be detected as individual objects because they emit up to 15\% of their light at [OIII] 5007\AA \ \citep{Dopita:1992,Schonberner:2010}. Moreover, their spatial distribution follows the stars, so their measured kinematics can be combined with integrated-light spectroscopy from the inner regions \citep{Coccato:2009}. Using current 8m-class telescopes, it is possible to generate large enough catalogs of planetary nebula velocities to constrain dynamical models of nearby galaxies at halo radii. 

We present a planetary-nebula survey of the galaxy NGC 1316. It was selected as a nearby, merger-remnant archetype. NGC 1316 is a radio source (Fornax A) with some similarities \citep{Schweizer:1980} to NGC 5128 (Cen A), the subject of one of the earliest and most successful planetary-nebula surveys \citep{Hui:1995}. Previous work  on NGC 1316 includes a survey \citep{Arnaboldi:1998} which detected 43 PNe. It is the brightest cluster galaxy (BCG) in Fornax and the dominant elliptical in the infalling subcluster described by \citet{Drinkwater:2001}.  It is assumed to be the remnant of a merger of two disk galaxies based on the intermediate age globular cluster system \citep{Goudfrooij:2001} and its dusty, unsettled appearance.

We present PN observations of the galaxy in Sect. \ref{Observations}. Sect. \ref{PNcats} covers the creation of the PN catalog including the separation of the overlapping NGC 1316 and NGC 1317 populations.  The results are presented in Sect. \ref{Results}, and our analysis, discussion of the implications and future work are are found in Sect. \ref{Discussion}. Throughout this work, we adopt the measured distance of 21.5Mpc for NGC 1316 \citep{Tonry:2001}, so $1^{\prime}=6.25$ kpc.

%%%%%%%%%%%%%%%%%%%%%%%%%%%%%%%%%%%%%%%%%%%%%%%%
%OBSERVATIONS
%%%%%%%%%%%%%%%%%%%%%%%%%%%%%%%%%%%%%%%%%%%%%%%%
\section{Planetary nebula observations}
\label{Observations}
\begin{figure}
\includegraphics[width=0.45\textwidth, angle=0]{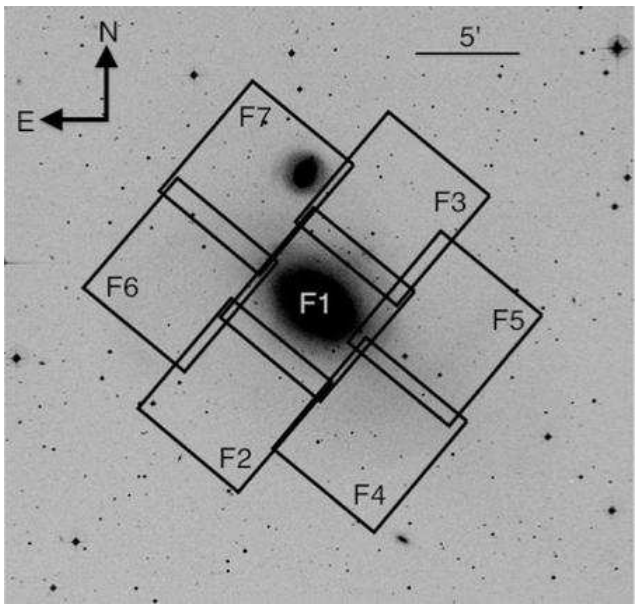}
\caption{ \label{1316dss} DSS image of NGC 1316 showing the approximate positions of the seven fields observed. NGC 1316 sits at the center of Field 1 and NGC 1317 is visible in Field 7. The dispersion direction is oriented along the major axis of NGC 1316. The actual observed area is slightly smaller because the two counter-dispersed exposures of each field do not overlap completely. }
\end{figure}

The planetary nebula kinematics  were measured using counter-dispersed spectroscopy with FORS2 on the VLT \citep{Appenzeller:1998}. As shown in Fig. \ref{1316dss}, NGC 1316 was observed with 7 fields on 3-5 November 2002. The observing conditions are summarized in Table \ref{1316olog}.

\begin{table}
\caption{Observing log for NGC 1316$^*$. }
\label{1316olog}
\centering
\begin{tabular}{c l l c c}
\hline \hline
 Field & P.A. Label & Exp. Time & Seeing\\
 && (s) & (FWHM)\\
\hline
NGC1316-1 & W140 & 4 $\times$ 1080 & $1.16^{\prime\prime}$\\
NGC1316-1 & E320 & 4 $\times$ 1080 & $1.26^{\prime\prime}$\\
NGC1316-2 & W140 & 4 $\times$ 1080 & $1.01^{\prime\prime}$\\
NGC1316-2 & E320 & 4 $\times$ 1080 & $1.13^{\prime\prime}$\\
NGC1316-3 & W140 & 4 $\times$ 1080 & $0.98^{\prime\prime}$\\
NGC1316-3 & E320 & 4 $\times$ 1080 & $1.16^{\prime\prime}$\\
NGC1316-4 & W140 & 4 $\times$ 1080 & $0.93^{\prime\prime}$\\
NGC1316-4 & E320 & 4 $\times$ 1080 & $0.95^{\prime\prime}$\\
NGC1316-5 & W140 & 4 $\times$ 1080 & $1.10^{\prime\prime}$\\
NGC1316-5 & E320 & 4 $\times$ 1080 & $1.10^{\prime\prime}$\\
NGC1316-6 & W140 & 4 $\times$ 1080 & $1.16^{\prime\prime}$\\
NGC1316-6 & E320 & 4 $\times$ 1080 & $1.11^{\prime\prime}$\\
NGC1316-7 & W140 & 4 $\times$ 1080 & $0.81^{\prime\prime}$\\
NGC1316-7 & E320 & 4 $\times$ 1080 & $0.92^{\prime\prime}$\\
\hline
\end{tabular}
\begin{list}{}{}
\item[$^{\mathrm{*}}$]  Each field is observed at P.A.=140$^{\circ}$ and 320$^{\circ}$, \ labeled W140 and E230 respectively.
\end{list}
\end{table}

The FORS field-of-view is 6$^{\prime}.8\times6^{\prime}.8$ or 42.5kpc on each side at 21.5 Mpc \citep{Tonry:2001}. The spatial scale is 0$^{\prime\prime}$.25 /pixel.

We used the 1400V  grating with a measured mean dispersion of 0.63\AA pix$^{-1}$. The measured FWHM of the arclines in the calibration frame was 3.8 pixels. 

An interference filter, FILT\_503\_5$+$86, was used to select emission at the wavelength of the redshifted [OIII] line. The filter prevents spectra of continuum sources from overwhelming the field, and it allows $>50$\%-transmission detection of PNe in the range of -720 to 2280 km s$^{-1}$.  A plot of the filter bandpass and a discussion of how well it is approximated by a Gaussian is available in Sect. 4.4.2 of \cite{McNeil:2010}.

In counter-dispersed spectroscopy, the field is imaged with a grism in the light path. This causes continuum sources such as stars to appear as streaks in the direction of dispersion. Emission-line objects such as planetary nebulae appear as dots. Another exposure is taken with the position angle (P.A.) rotated 180$^{\circ}$. The distance between the emission line objects detected in the P.A.=0 and P.A.=180 frames is a function of the object's velocity.

The direction of dispersion is along the horizontal axis ($x$-axis). Both the $x$- and $y$-axes contain spatial information due to the slitless technique. The grism causes an anamorphic distortion in the focal plane including contraction in the direction of dispersion.   

Although the reduction and calibration techniques used for this data are detailed in Sect. 2.3 of \citet{McNeil:2010}, we summarize them briefly here. The geometry of the field of view is mapped using a set of three daytime calibration frames based on a drilled mask with a grid of holes. The mask is imaged with white light in the first calibration frame. In the second and third frames, the light from the mask is dispersed using white and HeAr lamps respectively. In combination, the three frames allow us to measure the local dispersion, the local bandpass (which shifts across the field due to the converging beam and the interference filter) and the anamorphic distortion.

%%%%%%%%%%%%%%%%%%%%%%%%%%%%%%%%%%%%%%%%%%%%%%%%
%CATALOGS OF PNe
%%%%%%%%%%%%%%%%%%%%%%%%%%%%%%%%%%%%%%%%%%%%%%%%

\section{Cataloging and isolating the planetary nebulae}
\label{PNcats}
The observations were reduced and registered using similar techniques to those detailed in \citet{McNeil:2010}. Using registered images, PNe were detected by blinking the two counter-dispersed exposures. A detection is comprised of a pair of [OIII] 5007\AA \ emission-line sources (one in each exposure) of approximately the same brightness and at the same y-position. The separation between the objects must be consistent with the transmission limits of the filter. To estimate and improve the completeness of the catalog, two independent attempts at the detections are undertaken a few days or weeks apart. 

The observations for NGC 1316 were done using FORS2 instead of FORS1. This enabled a more effective calibration set-up with a drilled mask instead of the series of calibration slitlets used by \cite{McNeil:2010}. The distortion correction and registration is the same as the one described in \S 2.3 of \cite{McNeil:2010}, but the calibration positions appear less extended in the vertical direction, which makes them easier to measure.

The astrometry was performed following the steps described in \cite{McNeil:2010}. The position of each PNe is taken to be the mean of the detections in the two counter-dispersed exposures. The astrometric solution is based on the measured positions of the stars and the coordinates of the corresponding objects in the USNO-B catalog \citep{Monet:2003}. Our measured uncertainty (rms error of the solution) is about 0.\arcsec3, which includes the difficulty of measuring bandpass-shifted continuum sources.

\subsection{Velocity calibration}
The drilled-mask calibrations used in these observations are more accurate than the slitlets used for FORS1 data. There is no reason to suspect that the NGC 1316 PN velocities are incorrect in either a relative or an absolute sense. 

To check the agreement between the 7 fields, we matched duplicate detections in the areas of overlap. There were 78 duplicated PNe concentrated in fields that were vertically adjacent (Fields 1 \& 2, Fields 1 \& 3, Fields 4 \& 5, and Fields 6 \& 7). In these four cases, the measured shift between the fields was of the same order as the error in the shift, so it was not applied. The rms of the differences in the pairs was 42 km s$^{-1}$. We estimate the instrumental error of this sample to be 42/$\sqrt{2}\approx $30 km s$^{-1}$. 

\begin{figure}
\includegraphics[width=0.45\textwidth, angle=0]{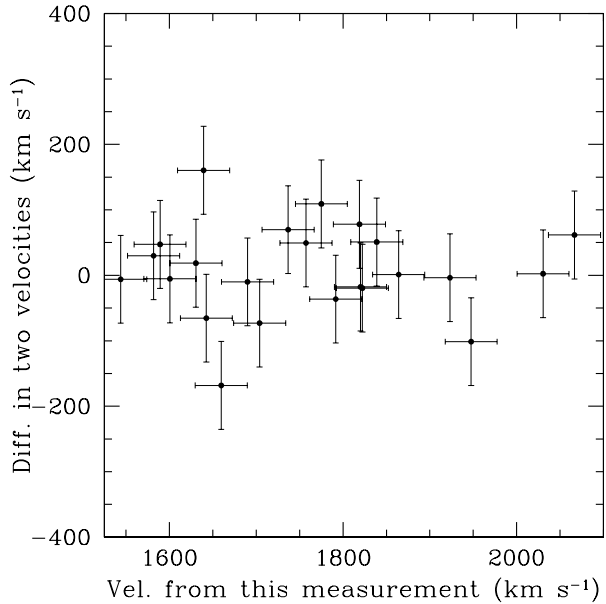}
\caption{ \label{1316MagdaCheck} Comparison of the velocities in this catalog and the sample from \cite{Arnaboldi:1998} for the 23 PNe in common. The previous measurement has been subtracted from the newer one on the vertical axis and the errors have been combined in quadrature. The median difference is 1.08 km s$^{-1}$. The scatter in the differences (70.7 km s$^{-1}$) is explained by the combined instrumental errors of the two samples. This indicates that the absolute velocity calibration of this sample is accurate. }
\end{figure}

We compared the measured PN velocities from this sample to those previously measured by \cite{Arnaboldi:1998}. 23 PNe are in common between the two samples. The median difference between the two measurements is 1.08km s$^{-1}$. The scatter in the difference is 70.7 km s$^{-1}$, which is roughly expected from the combination of the instrumental errors of each sample (30 km s$^{-1}$ and 60 km s$^{-1}$). The differences are plotted in Fig. \ref{1316MagdaCheck}. We conclude that our absolute velocity calibration is reliable. 

\subsection{Decomposing the NGC 1316 and NGC 1317 populations} \label{decompose}The NGC 1316 PN sample shows localized contamination from the nearby Fornax galaxy NGC 1317. It is an SABa galaxy almost directly North of NGC 1316. Their projected separation is 377\arcsec, and their systemic velocities are only 181km s$^{-1}$ apart (NGC 1316 at 1760 km s$^{-1}$ from Longhetti et al. 1998 and NGC 1317 at 1941 kms $^{-1}$ from deVaucouleurs et al. 1991). Due to NGC 1316's rotation, the velocity separation near NGC 1317 is even less. Although it is a much smaller galaxy, we expect there to be contamination from NGC 1317 in the field that covers its nucleus, Field 7. The decomposition was applied to Field 7 alone because it is the only field significantly influenced by NGC 1317.  The decomposition of the NGC 1316 and NGC 1317 PN samples follows a similar two-fold membership-assignment technique to the one described in \citet{McNeil:2010} for NGC 1399 and NGC 1404. Each PN is assigned a probability of membership in NGC 1316 based on its velocity and the surface brightness of the galaxy at that position. The probabilities are normalized such that all objects in the catalog belong to either NGC 1316 or NGC 1317, but membership is assigned only if there is greater than 90\% confidence. 

\subsubsection{Velocity criterion} 
\begin{figure}
\includegraphics[width=0.45\textwidth, angle=0]{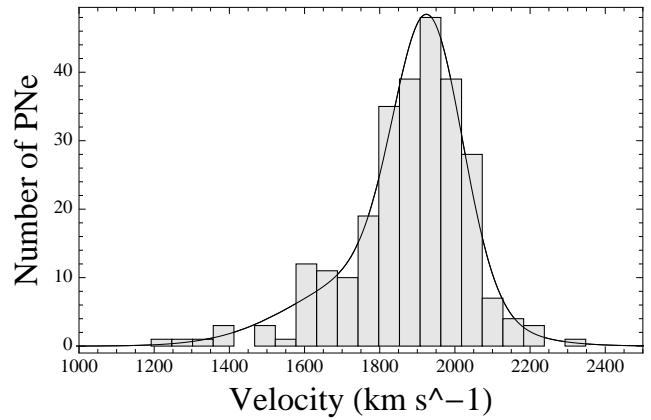}
\caption{ \label{decomphisto} Velocity distribution and histogram of the PNe from the F7 observations. The velocities are represented in the histogram. The smooth function represents the maximum-likelihood fit to the distribution--- it is parametrized as the sum of two Gaussians multiplied by a Gaussian representing the filter.}
\end{figure}

The first criterion for membership is based on velocity. Because the velocity separation is so small, and NGC 1317's systemic velocity is so near the edge of the filter, we considered the shape of the filter to be significant in shaping the overall velocity distribution. Using a maximum-likelihood technique, we fit the Field 7 velocity distribution with a sum of two Gaussians (representing the two galaxies) multiplied by a Gaussian (representing the filter). The fit to the velocity distribution is shown in Fig. \ref{decomphisto}. The fitted means of the two galaxies are found at 1826.86km s$^{-1}$ and 1936.08 km s$^{-1}$ compared to the observed values of 1760 km s$^{-1}$ and 1941km s$^{-1}$ \citep{Longhetti:1998, de-vaucouleurs:1991}. Accounting for rotation in the outer parts of NGC 1316, we find this fit is consistent with our physical understanding of the system. 

We then reverse the technique to calculate the probability of membership of each velocity in the NGC 1316 Gaussian multiplied by the filter versus the NGC 1317 Gaussian multiplied by the filter. The probabilities were normalized such that each PNe must either be a member of one galaxy or the other. The probability of membership in NGC 1316 defines our velocity membership criterion. 

\subsubsection{Spatial criterion} For the spatial criterion, we project our PNe onto the major axes of the two galaxies and estimate the light contributions from each galaxy at the location of each PN. We can use this selection method because PNe trace the stellar light \citep{Coccato:2009}. 

We estimate the major-axis radius of each PN assuming a fixed ellipticity of $\epsilon=0.33$ based on isophotes from the DSS image of NGC 1316 and $\epsilon=0.17$ for NGC 1317 based on measurements by \citet{Wozniak:1995}. The position angle of the photometric major axis of NGC 1316 is assumed to be constant at 50$^{\circ}$.  For NGC 1317, we estimate P.A.=150$^{\circ}$ based on Fig. 2 in \citet{Wozniak:1995}, but this is only valid because we are interested in values outside $\sim16$\arcsec.

We compared the light contribution from each galaxy by using the B-band light profiles from \citet{Schweizer:1980} for NGC 1316 and \citet{Wozniak:1995} for NGC 1317. To estimate the surface brightness profile beyond the limits of these observations, we fit the outermost 14 observations from Schweizer with a Sersic profile and the outermost 15 observations from Wozniak et al. with an exponential profile. The probability of membership based on position is the ratio of the surface brightness of NGC 1316 to the surface brightness of NGC 1317. 

\subsubsection{Selection of the NGC 1316 sample}
To select the final sample for NGC 1316, we multiplied the velocity and spatial criteria. The final probabilities are normalized so that the total probability of membership for each object is one. However, objects are only assigned membership in a galaxy if they have more than 90\% probability of belonging. This yields a catalog of 796 PNe bound to NGC 1316. 69 PNe are members of NGC 1317 and 43 are ambiguous. 

\subsection{Contamination from other sources}
Based on the two-fold decomposition technique in Sect. \ref{decompose}, we are able to identify and remove PNe that belong to nearby galaxies. Our observations also include possible contamination from other sources. Most background galaxies can be identified as non-planetary nebulae because they are associated with continuum, they are spatially extended or they have double emission lines. These contaminants are removed before the sample is assembled. 

However, some emission-line sources cannot be distinguished from PNe---they are most likely unresolved Lyman-$\alpha$ galaxies at $z=3.1$. These background objects show a velocity distribution centered on the filter instead of the galaxy's systemic velocity enabling us to estimate how much contamination is present in the catalog. In the case of NGC 1316, these objects would have lower velocities on average than the PNe from the galaxy. However, they are extremely difficult to remove because their spatial distribution is approximately flat over the field of view. Contamination from these background galaxies is minimized by avoiding faint objects in the selection process because their equivalent widths are low and they are less likely to be continuum-free \citep{Arnaboldi:2002}. In the case of the NGC 1316 sample, contamination is limited (see the excess in the histogram at low velocities in Fig. \ref{1316velhisto}) and we are able to effectively analyze our sample with methods that are robust to outliers.

\subsection{PN catalogs}
\subsubsection{Finalizing the catalogs}

\begin{figure}
\includegraphics[width=0.45\textwidth]{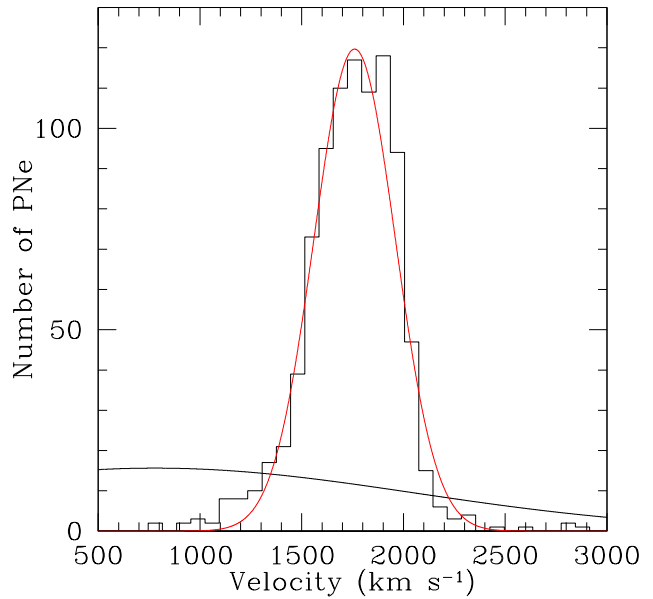}
\caption{ \label{1316velhisto} Velocity histogram of the PNe detected in the NGC 1316 observations. This includes objects associated with NGC 1316 and NGC 1317. The red Gaussian represents NGC 1316 and the black curve indicates the transmission of the filter in the relevant velocity range. Some low-velocity objects are skewing the distribution, and we see the suggestion of a second peak near the systemic velocity of NGC 1317. (1941 km s$^{-1}$ according to de Vaucouleurs et al. 1991) }
\end{figure}

%Emily: See publish1316.dat for the full table to send to CDS.
\begin{table}
\caption{Catalog of PNe in NGC 1316}
\label{1316table}
\begin{center}
\setlength\tabcolsep{5pt}
\begin{tabular}{cllc l}
\hline \hline
\noalign{\smallskip}
   ID         &   RA       &  Dec          &    wavelength     &    $v_{\rm helio}$   \\
      &   J2000    &  J2000        &    \AA            &    km s$^{-1}$   \\
\hline\\      
 NGC1316   1    &    50.47341    &    -37.21804      &       5039.0        &        1924    \\        	
 NGC1316   2    &    50.47824    &    -37.23466      &       5033.4        &        1587    \\        	
 NGC1316   3    &    50.47914    &    -37.21044      &       5038.7        &        1901    \\        	
 NGC1316   4    &    50.48403    &    -37.22564      &       5035.3        &        1700    \\        	
 NGC1316   5    &    50.48860    &    -37.24886      &       5031.5        &        1472    \\        	
 NGC1316   6    &    50.48942    &    -37.21301      &       5036.9        &        1798    \\        	
 NGC1316   7    &    50.49080    &    -37.24301      &       5036.4        &        1764    \\        	
 NGC1316   8    &    50.49086    &    -37.20056      &       5033.6        &        1598    \\        	
 NGC1316   9    &    50.49483    &    -37.20033      &       5035.5        &        1713    \\        	
 NGC1316  10    &    50.49716    &    -37.23789      &       5033.2        &        1572    \\  
\multicolumn{1}{c}{$\vdots$}    &
\multicolumn{1}{c}{$\vdots$}    &
\multicolumn{1}{c}{$\vdots$}    &
\multicolumn{1}{c}{$\vdots$}    &
\multicolumn{1}{c}{$\vdots$}    \\
 NGC1316 794    &    50.86464    &    -37.18436      &       5038.1        &        1868    \\        	
 NGC1316 795    &    50.86579    &    -37.18282      &       5039.4        &        1947    \\        	
 NGC1316 796    &    50.87387    &    -37.21063      &       5038.2        &        1876    \\   
\hline\\
\end{tabular}\label{tab:cat}
\end{center}
\begin{minipage}{8 cm}
NOTES - Details on the astrometry are in Sect. \ref{PNcats}. Velocities, measured from the [OIII] 5007\AA \ line, are corrected to heliocentric. The velocity error is 30 km s$^{-1}$. The full catalog is available in the associated electronic material.
\end{minipage}
\end{table}

%Emily: See publish1317.dat for the full table to send to CDS.
\begin{table}
\caption{Catalog of PNe in NGC 1317}
\label{1317table}
\begin{center}
\setlength\tabcolsep{5pt}
\begin{tabular}{cllc l}
\hline \hline
\noalign{\smallskip}
   ID         &   RA       &  Dec          &    wavelength     &    $v_{\rm helio}$   \\
      &   J2000    &  J2000        &    \AA            &    km s$^{-1}$   \\
\hline\\      
   NGC1317    1   	&     50.65941 	&   -37.09871   	&     5039.8   	& 	  1970		\\ 	
   NGC1317    2   	&     50.66192 	&   -37.09394   	&     5040.7   	& 	  2026		\\ 	
   NGC1317    3   	&     50.66277 	&   -37.10502   	&     5040.3   	& 	  1998		 \\	
   NGC1317    4   	&     50.66303 	&   -37.10728   	&     5041.5   	& 	  2072		\\ 	
   NGC1317    5   	&     50.66341 	&   -37.10949   	&     5040.1   	& 	  1989		 \\	
   NGC1317    6   	&     50.66537 	&   -37.08525   	&     5039.6   	& 	  1956		 \\	
   NGC1317    7   	&     50.66625 	&   -37.09976   	&     5041.6   	& 	  2081		 \\	
   NGC1317    8   	&     50.66925 	&   -37.10303   	&     5039.5   	& 	  1951		\\ 	
   NGC1317    9   	&     50.66943 	&   -37.08724   	&     5039.1   	& 	  1927		 \\	
   NGC1317   10   	&     50.66995 	&   -37.10643   	&     5039.5   	& 	  1954		\\ 
  \multicolumn{1}{c}{$\vdots$}    &
\multicolumn{1}{c}{$\vdots$}    &
\multicolumn{1}{c}{$\vdots$}    &
\multicolumn{1}{c}{$\vdots$}    &
\multicolumn{1}{c}{$\vdots$}    \\
   NGC1317   67   	&     50.71118 	&   -37.09235   	&     5039.2   	& 	  1933		  \\	
   NGC1317   68   	&     50.71183 	&   -37.09738   	&     5039.3   	& 	  1938		  \\	
   NGC1317   69   	&     50.72124 	&   -37.10676   	&     5039.1   	& 	  1926		  \\
\hline\\
\end{tabular}\label{tab:cat}
\end{center}
\begin{minipage}{8 cm}
NOTES - Details on the astrometry are in Sect. \ref{PNcats}. Velocities, measured from the [OIII] 5007\AA \ line, are corrected to heliocentric. The velocity error is 30 km s$^{-1}$. The full catalog is available in the associated electronic material.
\end{minipage}
\end{table}

The velocity histogram of 908 PNe observed in the NGC 1316 fields is shown in Fig. \ref{1316velhisto}. This is an exceptionally large catalog. 796 of these objects are bound to NGC 1316 with greater than 90\% probability. The sample also includes 69 members of NGC 1317's PN population and 43 ambiguous objects. The catalogs for NGC 1316 and NGC 1317 are presented in Tables \ref{1316table} and \ref{1317table} respectively.

\subsubsection{Comparing the efficiency of our detections}
The expected number of detections for our sample is determined by the observations and the intrinsic properties of the galaxies. We can judge the efficiency of each observing run by scaling the size of the samples by these factors. 

This sample is comprised of nearly 5.5 times as many objects as the NGC 1399 catalog presented in \cite{McNeil:2010}. Since they are both members of the Fornax cluster, we assume a common distance and compute the size of a hypothetical sample for comparison. 

Beginning with the 146 PNe detected in NGC 1399, we add 55 objects based on symmetry arguments to account for the 4 of 9 fields that were not observed. We scale by 2.42 to account for the difference in brightness. For this we used 2MASS \citep{Skrutskie:2006} K magnitudes adjusted according to \cite{Kormendy:2011}. Finally, we consider the relation from \cite{Buzzoni:2006} relating the PN number density to the color of the galaxies. The overall trend implies that redder galaxies have a higher specific frequency. Despite being reddened by dust, NGC 1316 is bluer than NGC 1399 implying an increase in our predicted sample. There is significant scatter in the Buzzoni et al. relation, accounted for by variation in the lifetimes of PN populations, and a much steeper relation for external ellipticals than for Local Group galaxies. The adjustment for our predicted sample is between 1.2 and 2.5. 

Adjusting for these differences, we can compare our sample of 796 PNe in NGC 1316 to a predicted sample between 588 and 1216 PNe. We conclude that, although NGC 1316 has a remarkably large sample, it is not unexpected.

%%%%%%%%%%%%%%%%%%%%%%%%%%%%%%%%%%%%%%%%%%%%%%%%
%RESULTS
%%%%%%%%%%%%%%%%%%%%%%%%%%%%%%%%%%%%%%%%%%%%%%%%

\section{Results}
\label{Results}
\subsection{The spatial distribution}

\begin{figure}
\includegraphics[width=0.45\textwidth]{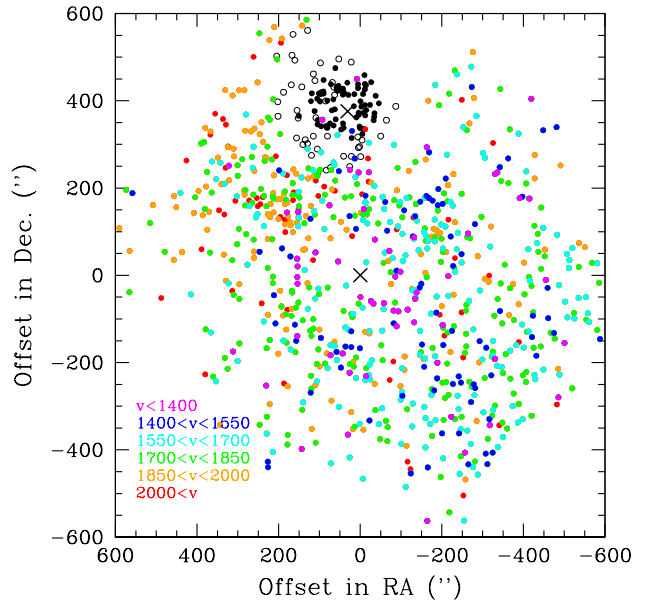}
\caption{ \label{1316space} Spatial distribution of the 908 objects in the NGC 1316 observations. North is up and East is left. The centers of the two galaxies are marked with black crosses. Surrounding NGC 1317, the solid, black dots represent PNe bound to NGC 1317 with $>$90\%probability. The colored dots indicate membership in NGC 1316. If a PN cannot be assigned membership with $>$90\% probability, it is marked as ambiguous (hollow black dots). Ignoring the structure from this companion galaxy, this figure shows NGC 1316's rotation oriented along P.A.=64$^{\circ}$ (see Sect. \ref{rotation}). The north-east is rotating away from the observer while the south-west is moving towards us relative to the mean.   }
\end{figure}
The spatial distribution of the PNe in NGC 1316 is shown in Fig. \ref{1316space}. The overall shape is determined by the layout of the 7 observed fields in Fig. \ref{1316dss}, and we see rotation along the major axis (P.A.=64$^{\circ}$, see Sect. \ref{rotation}).

\begin{figure}
\includegraphics[width=0.45\textwidth]{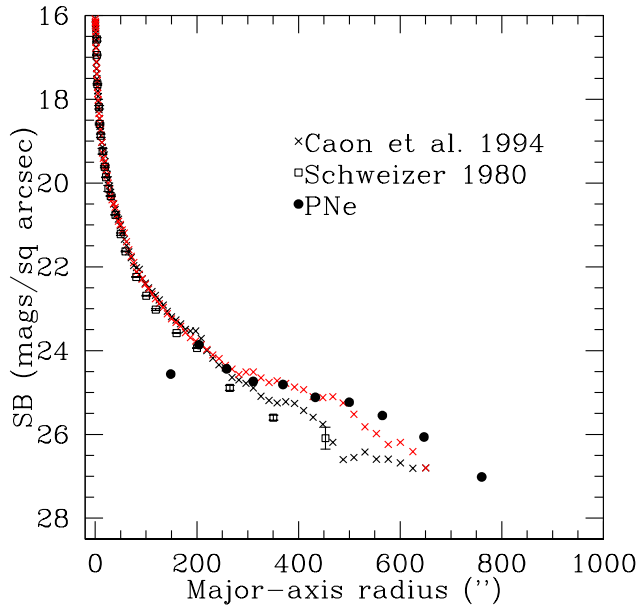}
\caption{ \label{1316numdens} Surface density profiles for the PNe and stars in NGC 1316. Red and black crosses mark the NE and SW major-axis B-band surface brightness profiles respectively \citep{Caon:1994}. The hollow squares show the measurements from \cite{Schweizer:1980}. The agreement with the NE major axis between 200$^{\prime\prime}$ and 500 $^{\prime\prime}$ indicates that the PNe do trace the light, except where dust obscures the observations. }
\end{figure}

The analysis of the PN kinematics and their use as tracers in dynamical models are dependent upon the consistency of PNe and stars. \cite{Coccato:2009} showed this for a sample of elliptical galaxies observed with the Planetary Nebulae Spectrograph.  In Fig. \ref{1316numdens}, we show a comparison of the stellar surface brightness and the PN number density. The PNe were divided into annular bins and the logarithmic projected number density in each bin was calculated as 
 \[
	\mu_{\mathrm{PNe}}=-2.5 \log(N_{\mathrm{PNe}}/A) + c,
\]
where $N_{\rm{PNe}}$ is the number of PNe in the bin, $A$ is the area of the annulus and $c$ is an arbitrary constant selected for the galaxy (but remaining constant for all the bins). 

The first bin is incomplete because of the void in the center where PNe are not detectable. The surface brightness profiles are in good agreement until $\sim250^{\prime\prime}$ where the influence of the dust becomes apparent. Dust strongly affects the integrated light observations in the outer parts. The NE major axis is the least affected, and, therefore, is the best point of comparison for the PNe. 

In the region where the PN sample is complete and dust does not obscure the B-band integrated light, the PN number density shadows the NE major-axis surface density of the stars. We conclude that the PNe and the integrated light are compatible tracers of the underlying stellar mass. We will use the two tracers together to constrain dynamical models of NGC 1316 in \S \ref{DynModels}. 

\subsection{The kinematics}
\label{rotation}
\begin{figure}
\includegraphics[width=0.45\textwidth]{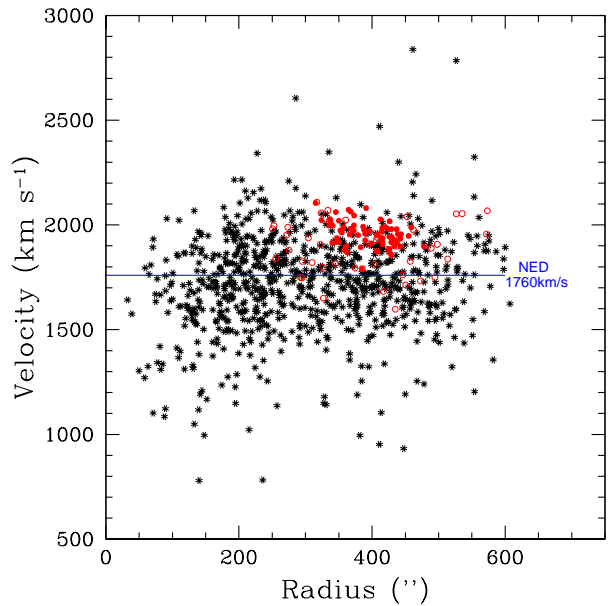}
\caption{ \label{1316phase} Line-of-sight velocity against projected radius. The sample is centered on the systemic velocity of 1760km/s \citep{Longhetti:1998}. Solid red objects have been assigned membership in NGC 1317 with $>$90\% probability. The hollow red dots are ambiguous-- they could not be assigned to either galaxy with more than $>$90\% probability. There are some extreme-velocity objects near the center of the filter (779km s$^{-1}$) and around 2700 km s$^{-1}$ that may be background galaxies.}
\end{figure}

A phase-space projection of line-of-sight velocity against radius is shown in Fig.\ref{1316phase}. There are some high- and low-velocity objects that may be contaminants, but they will not affect the robust dispersion measurements discussed in Sect. \ref{sigma}.

\begin{figure}
\includegraphics[width=0.45\textwidth]{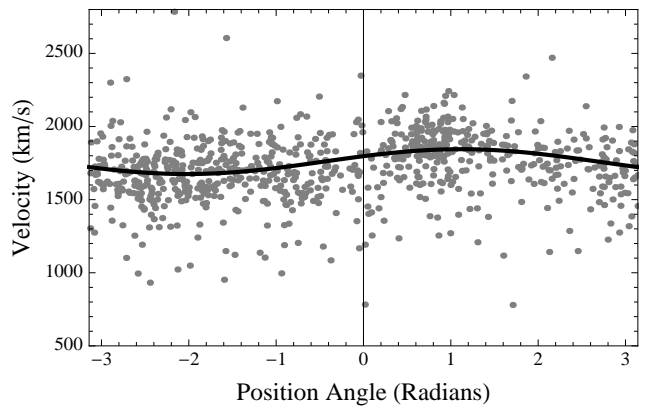}
\caption{ \label{1316rot} The line-of-sight velocities of the NGC 1316 PNe as a function of position angle. The grey dots represent the decomposed PNe sample. The black fit represents rotation of a spherically symmetric system with the axis of rotation in the plane of the sky. It indicates an amplitude of 85km s$^{-1}$ and a kinematic major axis at P.A.=64$^{\circ}$ (1.11 radians). }
\end{figure}

We quantify the rotation of the sample by modeling the total rotation and the rotation in annular bins.  S0 galaxies usually display rotation, and we are interested to learn to what extent the flattening is rotationally supported. The fitted rotation for the full sample of PNe is shown in Fig.\ref{1316rot}. 

For the fit, we assume that the galaxy is spherically symmetric, the amplitude of rotation is only a function of radius and that the axis of rotation is in the plane of the sky \citep{Cote:2001, Woodley:2010}. The velocity as a function of position angle, $\Theta$, is parametrized as 
\[
	v_p[\Theta]=V_{\rm{sys}}+ A \cos(\Theta-\Theta_0),
\]
where $V_{\rm{sys}}$ is the systemic velocity, $A$ is the amplitude of rotation, and $\Theta_0$ is the P.A. of the kinematic major axis. We also considered adding a $\sin (3\Theta)$ term, but the simpler model was preferred by a Bayesian Information Criterion test \citep{Schwarz:1978}.

Because we suspect a fit including a free systemic velocity to be influenced by the contamination from the center of the filter, we hold $V_{\rm{sys}}$ constant at 1760 km s$^{-1}$ \citep{Longhetti:1998}. The result is a kinematic major axis at P.A.=64$\pm8^{\circ}$ and an amplitude of rotation of 85$\pm11$ kms$^{-1}$. The photometric major axis is at P.A.=50$^{\circ}$.

\begin{figure}
\includegraphics[width=0.45\textwidth]{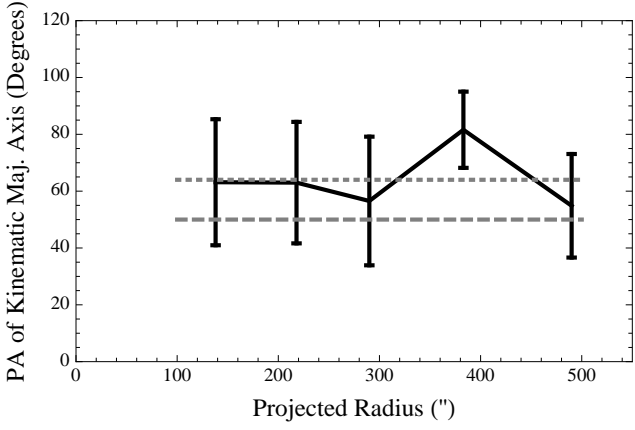}
\caption{ \label{1316PA} Fitted values for the position angle of the kinematic major axis of NGC 1316. The connected black points show the kinematic major axis's P.A. known from fitting the rotation in five radial bins. The P.A. fitted to the entire sample is 64$\pm$8$^{\circ}$ (shown as grey dotted line). The dashed grey line shows the photometric major axis at 50$^{\circ}$. }
\end{figure}

\begin{figure}
\includegraphics[width=0.45\textwidth]{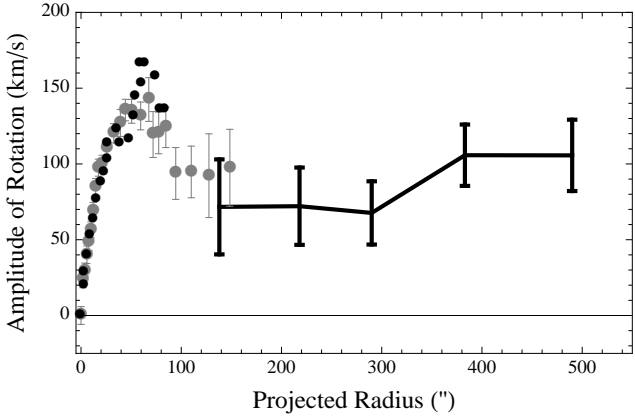}
\caption{ \label{1316rotcurve} The rotation curve of the PNe in the outer parts of NGC 1316. The black line traces the amplitude of rotation of the PNe in annular bins. At small radii, the gray dots show positive major-axis rotation measurements from \cite{Bedregal:2006}, and the black dots show major-axis rotation measurements from \cite{Arnaboldi:1998}.  }
\end{figure}

Using the rotation fitted to the entire sample as an initial guess, we divided the PNe into five radial bins. The P.A. and amplitude of rotation are modelled for each bin, and the results are in Fig. \ref{1316PA} and Fig. \ref{1316rotcurve} respectively. The position angle of the kinematic major axis is offset from the photometric major axis by 1.6$\sigma$, but still consistent (see Fig. \ref{1316PA}).  The amplitude of rotation rises in the outer parts. This could indicate that some PNe bound to NGC 1317 with a systemic velocity of 1941 km s$^{-1}$ \citep{de-vaucouleurs:1991} have stayed in our sample. This scenario is not supported by the distribution of the PNe in Fig. \ref{1316rot} nor the consistency of bin 5's position angle with the inner bins in Fig. \ref{1316PA}.  The rise could also be a real description of the kinematics of NGC 1316. This phenomenon could be evidence of the outward transfer of angular momentum resulting from a merger \citep{Quinn:1988}.

%%%%%%%%%%%%%%%%%%%%%%%%%%%%%%%%%%%%%%%%%%%%%%%%
%DISCUSSION
%%%%%%%%%%%%%%%%%%%%%%%%%%%%%%%%%%%%%%%%%%%%%%%%

\section{Discussion}
\label{Discussion}
\subsection{Line-of-sight velocity dispersion}
\label{sigma}
The line-of-sight velocity dispersion profile is a useful tool to illustrate the mass distribution and dynamics of elliptical galaxies. For our measurements, we use a robust technique for measuring the dispersion of discrete tracers that is described in \citet{McNeil:2010}. Because the calculation is only based on objects lying within 2$\sigma$ of the mean, outliers do not have a strong effect. This method is resilient against any lingering contamination from background galaxies. 

In the case of NGC 1316, we measure the dispersion after accounting for the rotation. The robust technique was applied to the residuals of the rotation fit in each of the five annular bins. 

\begin{figure}
\includegraphics[width=0.45\textwidth]{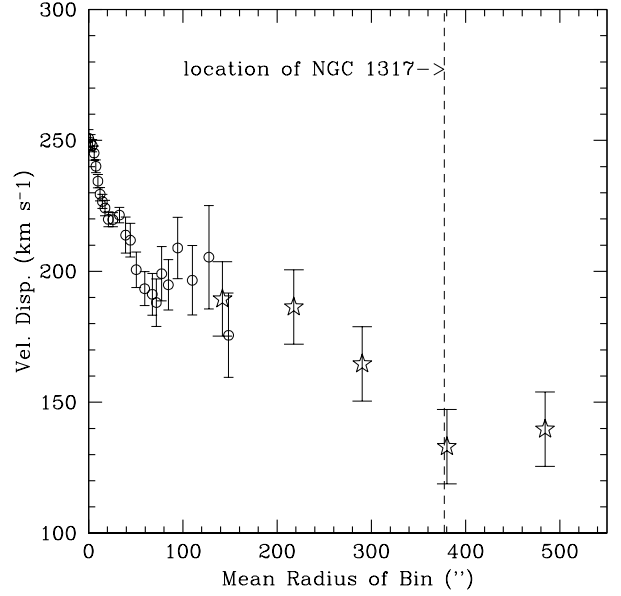}
\caption{ \label{1316sigma} Line-of-sight velocity dispersion profile for NGC 1316.  The hollow circles represent the integrated light measurements from \cite{Bedregal:2006} for the major axis. The PN dispersions (hollow stars) are computed using the robust technique on the residuals of the NGC 1316 rotation fit.  The instrumental dispersion is subtracted in quadrature. The results are resilient against outliers and describe the random motion in NGC 1316. The two tracers interface well at $\sim$150\arcsec. }
\end{figure}

The random-motion profile of NGC 1316 is shown in Fig. \ref{1316sigma}. At small radii, the integrated-light measurements from \cite{Bedregal:2006} show a falling dispersion profile.  As we move into the halo, the planetary nebulae indicate that the velocity dispersion continues to fall off until at least $\sim$ 500\arcsec.

\begin{figure}
\includegraphics[width=0.45\textwidth]{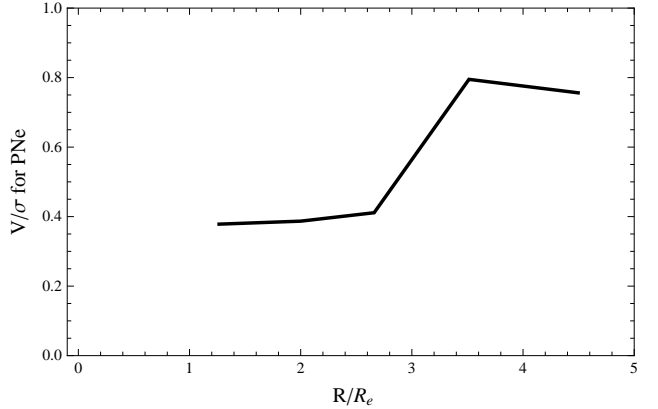}
\caption{ \label{1316vSigma} V/$\sigma$ profile of NGC 1316. The fitted rotation of the PNe is divided by the robust dispersion measurement shown in Fig. \ref{1316sigma} to illustrate the relative dynamical importance of rotation and random motions. For NGC 1316, rotation is more important in the outer parts than it is in the inner parts.  1R$_{\rm{e}}=109^{\prime\prime}$=11.4kpc \citep{Caon:1994}. }
\end{figure}

V/$\sigma$ is used to quantify the relative importance of rotation and random motions in the dynamics of the galaxy. We divide the amplitude of rotation  for the PNe by the robust dispersion shown in Fig. \ref{1316sigma}. The V/$\sigma$ profile for the planetary nebulae in NGC 1316 is shown in Fig. \ref{1316vSigma}. Rotation is more important in the outer parts than it is in the inner parts of NGC 1316, suggesting that angular momentum transferred outwards during the merger \citep{Quinn:1988}.

\subsection{Jeans Models}
\label{DynModels}
To better understand the meaning of the kinematics of the PN sample in NGC 1316, we construct a simple dynamical model of the galaxy. We begin with the Jeans equations \citep{Binney:1987}, assuming spherical symmetry: 
\begin{equation} \label{E:BT4-54}
	\frac{1}{\rho} \frac{d(\rho \sigma^2_r(r) )}{d r} + 2 \frac{\beta_0 \sigma^2_r(r) }{r} = - \frac{d\Phi}{d r},
\end{equation}
where $\rho$ is the spatial density, $\sigma_r$ is the radial component of the velocity dispersion, $\beta_0$ is the anisotropy of the velocity distribution ($\beta \equiv 1- \frac{\sigma^2_{\theta}}{\sigma^2_{\rm{r}}}$), which is assumed to be independent of radius here, and $\Phi$ is the total potential. The expressions on the left-hand side of Eq. \ref{E:BT4-54} can be related to observationally accessible quantities. In this case, we solve for the velocity dispersion and correct for the projection onto the plane of the sky. The projected velocity dispersion of the model is given as 
\begin{equation} \label{E:sigmaProj}
	\sigma^2_p(R) = \frac{2}{I(R)}  \int_R^{\infty}  \left( 1- \beta_0\frac{R^2}{r^2}\right) \frac{ j(r) \sigma_r^2(r)  r}{\sqrt{r^2-R^2}}  dr,
\end{equation}
where $I(R)$ is the observed surface brightness profile and j(r) is the light density which is calculated from the mass density, $\rho$, divided by a constant mass-to-light ratio, $\Gamma$.

The velocity dispersion of the model, $\sigma_r$ depends on the total mass of the system. As a first attempt, we assume a model where mass traces the light with a constant mass-to-light ratio.  We compared the model's projected velocity dispersion profile to the observations of integrated light \citep{Bedregal:2006} and the robust dispersion measurements of the PN sample. The integrated light positions were adjusted by -0.32kpc to maximize the symmetry of the dispersion profile\footnotemark. Since this non-rotating model does not otherwise account for the energy associated with rotation, we constrain our model with the observational quantity $\sigma_{tot}$. In the case of the integrated light, this is the line-of-sight velocity and the velocity dispersion added in quadrature. In the case of the PNe, it is the robust dispersion of the binned PNe velocities without adjusting for the rotation. The goodness-of-fit was quantified with a reduced $\chi^2$-style merit function
\begin{equation}
\label{merit function}
	\widetilde{\chi}^2=\frac{1}{N-n} \sum_{i=1}^{N} \frac{(\sigma_{\rm{tot}}(R_i) - \sigma_p(R_i))^2}{\sigma_{\rm{error}}^2(i)},
\end{equation}
where N is the number of data points constraining the fit and n is the number of free parameters, $\sigma_{\rm{tot}}$ is the observed value of sigma at a radius, $R_i$, $\sigma_p(R_i)$ is the projected dispersion of the model at $R_i$ and $\sigma_{\rm{error}}$ is the uncertainty of the observation belonging to that radius. 

\footnotetext{\cite{Bedregal:2006} selected the peak of the dispersion profile as their position zero-point. Upon closer inspection, the highest dispersion value was an anomaly in an otherwise symmetrical profile. We removed the anomalous point and fitted the profile with a symmetrical function finding a shift of -0.32 kpc relative to their system. The fit shared the same peak as the measurements (with the object representing the previous zero-point removed). }

\begin{figure}
\includegraphics[width=0.45\textwidth]{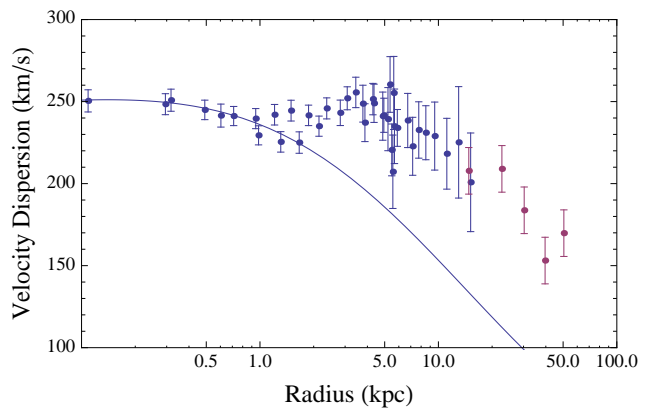}
\caption{ \label{1316SCmodel} Self-consistent Jeans model of NGC 1316. Blue points are integrated-light dispersions from \cite{Bedregal:2006} and the red points show the dispersions of PNe in annular bins.  The blue curve represents a model with a mass-to-light ratio of 4.0$M_{\odot}/L_{\odot}$ and a constant anisotropy of 0.44. The disagreement between the model and the observations in the outer part underlines the importance of DM in the dynamics of NGC 1316.  }
\end{figure}

The best fitting self-consistent model is shown in Fig. \ref{1316SCmodel}. We assume that the total mass traces the light with a constant mass-to-light ratio.  This model has a mass-to-light ratio of 4.0$M_{\odot}/L_{\odot}$ and an anisotropy of 0.44.
The disagreement between the model and the observations in the outer parts indicates the necessity of a dark matter component, in this case parameterized as a pseudo-isothermal distribution such that 
\begin{equation} 
\label{E:DMdist}
	\rho_{\mathrm{DM}}(r)=\frac{\sigma^{2}_{\mathrm{DM}}}{2\pi G (a^2 +r^2)},
	\end{equation}
where $\sigma_{\mathrm{DM}}$ is a velocity dispersion representing the total mass of the DM halo and $a$ is a core radius. 

Adding in the dark matter component, we explored a grid of models varying over the four parameters $\beta_0$, $\Gamma$, $\sigma_{\mathrm{DM}}$, and $a$. We resolved the anisotropy ($\beta$) to 0.01, the mass-to-light ratio ($\Gamma$) to 0.1$M_{\odot}/L_{\odot}$, the core radius ($a$) to 1 kpc and the dark matter dispersion ($\sigma_{\mathrm{DM}}$) to 1km s$^{-1}$. 

\begin{figure}
\includegraphics[width=0.45\textwidth]{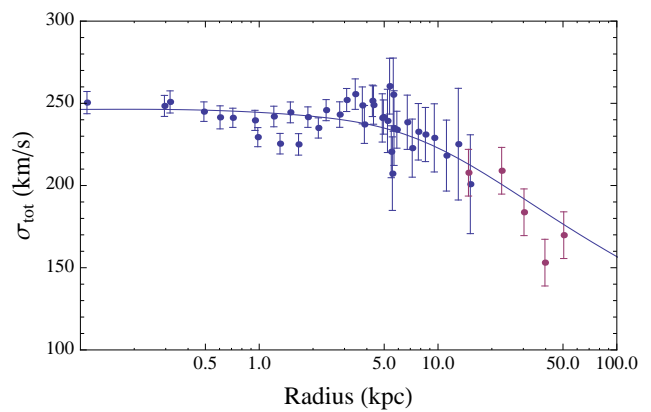}
\caption{ \label{1316DMmodel} Dynamical model of NGC 1316 including dark matter. The blue points are integrated-light dispersions from \cite{Bedregal:2006} and the red points are measured from PNe. The model (blue curve) has a constant mass-to-light ratio of 2.8 $M_{\odot}/L_{\odot}$ and a constant anisotropy of 0.50. The dark matter distribution is characterized by $\sigma_{\mathrm{DM}}=289$ km s$^{-1}$ and a core radius of 5 kpc. This model does not rotate. }
\end{figure}

The best value for our merit function corresponded to a model with $\beta_0=0.50$, $\Gamma=2.8{M}_{\odot}/{L}_{\odot}$, $a=5$kpc, and $\sigma_{\mathrm{DM}}=289$ km s$^{-1}$.  The best model is shown in Fig. \ref{1316DMmodel}.

The modeling of NGC 1316 is limited by the simplifications in the modeling technique. The outcome would be improved by assuming an axisymmetric distribution that is flattened by rotation. The mass-to-light ratio might well change over the radial range of our models. Allowing the anisotropy to change as a function of radius will add flexibility to the fitting and create a more accurate model of NGC 1316.

\begin{figure}
\includegraphics[width=0.45\textwidth]{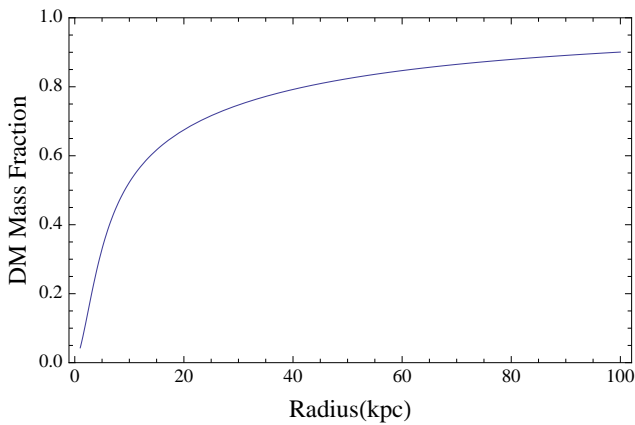}
\caption{ \label{1316DMMF} Dark matter mass fraction of NGC 1316. By considering the dark and luminous components our model separately, we can assess what fraction of the mass at each radius is DM. This plot is made for the model that best fits the kinematic data: $\beta_0=0.50$, $\Gamma=2.8M_{\odot}/{L}_{\odot}$, $a=5$kpc, and $\sigma_{\mathrm{DM}}=289$ km s$^{-1}$. }
\end{figure}

Despite the inherent assumptions in this model, it presents a realistic physical description that is consistent with observations and theory. Our mass-to-light ratio of 2.8 $M_{\odot}/L_{\odot}$ lands at the low end of values measured for other galaxies (see Fig 4.25 in \cite{Binney:2008}). The dark matter content of NGC 1316 shown in Fig. \ref{1316DMMF} is towards the high end of estimates of 5-65\% within 1R$_e$ \citep[$11.4\mathrm{kpc}=109^{\prime\prime}$,][]{Caon:1994}  \citep{Gerhard:2001, Treu:2004,Cappellari:2006,Thomas:2007, Barnabe:2009}. 

\subsection{Implications for the assembly of early-type galaxies}
Evidence from the globular cluster system \citep{Goudfrooij:2001} indicates that the last major merger occured 3 Gyr ago. The resulting galaxy is dusty with dynamically important rotation. Projecting the evolution of NGC 1316 forward in time, the dust will settle into a disk \citep[and references therein]{van-der-Kruit:2011}. A system like this could evolve into an object like the Sombrero Galaxy (NGC 4594, M104). 

NGC 1316 provides a lower limit for the time between a major merger and a galaxy like NGC 4594 to form. Likewise, it constrains the time since NGC 4594 last underwent a major merger to be greater than 3 Gyr. In the context of mass assembly within clusters, NGC 1316 tells a different story to the Fornax cD galaxy, NGC 1399. NGC 1399 is a slow rotating, possibly triaxial, dominant galaxy. The cD envelope implies that accretion is currently dominating its mass assembly. In contrast, NGC 1316 is in a phase where it has recently undergone a major merger and would need to be involved in several more to resemble NGC 1399. As shown in Fig. \ref{1316rotcurve} and \ref{1316sigma}, NGC 1316's rotation stays constant in the outer parts, but the random motions decrease.  Despite dominating its subcluster, NGC 1316 illustrates an earlier phase in mass assembly within clusters. 

\subsection{Future work}
The ultimate goal of these observations is to constrain the outer parts of dynamical models. We have presented spherical, non-rotating Jeans models. Future work will focus on more sophisticated models including NMAGIC modeling \citep[see][]{deLorenzi:2008}. 

\begin{acknowledgements} We thank A. Bedregal for generously sharing the major-axis kinematics of NGC 1316. We thank W. Harris for discussions regarding the rotation of NGC 1316.  This work has made use of the NASA/IPAC Extragalactic Database (NED) which is operated by the Jet Propulsion Laboratory, California Institute of Technology, under contract with the National Aeronautics and Space Administration. \end{acknowledgements}

\bibliography{/Users/emcneil/School/References/References}

\begin{thebibliography}{42}
\expandafter\ifx\csname natexlab\endcsname\relax\def\natexlab#1{#1}\fi

\bibitem[{{Appenzeller} {et~al.}(1998){Appenzeller}, {Fricke}, {F{\"u}rtig},
  {G{\"a}ssler}, {H{\"a}fner}, {Harke}, {Hess}, {Hummel}, {J{\"u}rgens},
  {Kudritzki}, {Mantel}, {Meisl}, {Muschielok}, {Nicklas}, {Rupprecht},
  {Seifert}, {Stahl}, {Szeifert}, \& {Tarantik}}]{Appenzeller:1998}
{Appenzeller}, I., {Fricke}, K., {F{\"u}rtig}, W., {et~al.} 1998, The
  Messenger, 94, 1

\bibitem[{{Arnaboldi} {et~al.}(2002){Arnaboldi}, {Aguerri}, {Napolitano},
  {Gerhard}, {Freeman}, {Feldmeier}, {Capaccioli}, {Kudritzki}, \&
  {M{\'e}ndez}}]{Arnaboldi:2002}
{Arnaboldi}, M., {Aguerri}, J.~A.~L., {Napolitano}, N.~R., {et~al.} 2002, \aj,
  123, 760

\bibitem[{{Arnaboldi} {et~al.}(1998){Arnaboldi}, {Freeman}, {Gerhard},
  {Matthias}, {Kudritzki}, {M{\'e}ndez}, {Capaccioli}, \&
  {Ford}}]{Arnaboldi:1998}
{Arnaboldi}, M., {Freeman}, K.~C., {Gerhard}, O., {et~al.} 1998, \apj, 507, 759

\bibitem[{{Barnab{\`e}} {et~al.}(2009){Barnab{\`e}}, {Czoske}, {Koopmans},
  {Treu}, {Bolton}, \& {Gavazzi}}]{Barnabe:2009}
{Barnab{\`e}}, M., {Czoske}, O., {Koopmans}, L.~V.~E., {et~al.} 2009, \mnras,
  399, 21

\bibitem[{{Bedregal} {et~al.}(2006){Bedregal}, {Arag{\'o}n-Salamanca},
  {Merrifield}, \& {Milvang-Jensen}}]{Bedregal:2006}
{Bedregal}, A.~G., {Arag{\'o}n-Salamanca}, A., {Merrifield}, M.~R., \&
  {Milvang-Jensen}, B. 2006, \mnras, 371, 1912

\bibitem[{{Binney} \& {Tremaine}(1987)}]{Binney:1987}
{Binney}, J. \& {Tremaine}, S. 1987, {Galactic Dynamics} (Princeton University
  Press)

\bibitem[{{Binney} \& {Tremaine}(2008)}]{Binney:2008}
{Binney}, J. \& {Tremaine}, S. 2008, {Galactic Dynamics: Second Edition}
  (Princeton University Press)

\bibitem[{{Buzzoni} {et~al.}(2006){Buzzoni}, {Arnaboldi}, \&
  {Corradi}}]{Buzzoni:2006}
{Buzzoni}, A., {Arnaboldi}, M., \& {Corradi}, R.~L.~M. 2006, \mnras, 368, 877

\bibitem[{{Caon} {et~al.}(1994){Caon}, {Capaccioli}, \&
  {D'Onofrio}}]{Caon:1994}
{Caon}, N., {Capaccioli}, M., \& {D'Onofrio}, M. 1994, \aaps, 106, 199

\bibitem[{{Cappellari} {et~al.}(2006){Cappellari}, {Bacon}, {Bureau}, {Damen},
  {Davies}, {de Zeeuw}, {Emsellem}, {Falc{\'o}n-Barroso}, {Krajnovi{\'c}},
  {Kuntschner}, {McDermid}, {Peletier}, {Sarzi}, {van den Bosch}, \& {van de
  Ven}}]{Cappellari:2006}
{Cappellari}, M., {Bacon}, R., {Bureau}, M., {et~al.} 2006, \mnras, 366, 1126

\bibitem[{{Coccato} {et~al.}(2009){Coccato}, {Gerhard}, {Arnaboldi}, {Das},
  {Douglas}, {Kuijken}, {Merrifield}, {Napolitano}, {Noordermeer},
  {Romanowsky}, {Capaccioli}, {Cortesi}, {de Lorenzi}, \&
  {Freeman}}]{Coccato:2009}
{Coccato}, L., {Gerhard}, O., {Arnaboldi}, M., {et~al.} 2009, \mnras, 394, 1249

\bibitem[{{C{\^o}t{\'e}} {et~al.}(2001){C{\^o}t{\'e}}, {McLaughlin}, {Hanes},
  {Bridges}, {Geisler}, {Merritt}, {Hesser}, {Harris}, \& {Lee}}]{Cote:2001}
{C{\^o}t{\'e}}, P., {McLaughlin}, D.~E., {Hanes}, D.~A., {et~al.} 2001, \apj,
  559, 828

\bibitem[{{de Lorenzi} {et~al.}(2008){de Lorenzi}, {Gerhard}, {Saglia},
  {Sambhus}, {Debattista}, {Pannella}, \& {M{\'e}ndez}}]{deLorenzi:2008}
{de Lorenzi}, F., {Gerhard}, O., {Saglia}, R.~P., {et~al.} 2008, \mnras, 385,
  1729

\bibitem[{{de Vaucouleurs} {et~al.}(1991){de Vaucouleurs}, {de Vaucouleurs},
  {Corwin}, {Buta}, {Paturel}, \& {Fouque}}]{de-vaucouleurs:1991}
{de Vaucouleurs}, G., {de Vaucouleurs}, A., {Corwin}, Jr., H.~G., {et~al.}
  1991, \skytel, 82, 621

\bibitem[{{Doherty} {et~al.}(2009){Doherty}, {Arnaboldi}, {Das}, {Gerhard},
  {Aguerri}, {Ciardullo}, {Feldmeier}, {Freeman}, {Jacoby}, \&
  {Murante}}]{Doherty:2009}
{Doherty}, M., {Arnaboldi}, M., {Das}, P., {et~al.} 2009, \aap, 502, 771

\bibitem[{{Dopita} {et~al.}(1992){Dopita}, {Jacoby}, \&
  {Vassiliadis}}]{Dopita:1992}
{Dopita}, M.~A., {Jacoby}, G.~H., \& {Vassiliadis}, E. 1992, \apj, 389, 27

\bibitem[{{Drinkwater} {et~al.}(2001){Drinkwater}, {Gregg}, \&
  {Colless}}]{Drinkwater:2001}
{Drinkwater}, M.~J., {Gregg}, M.~D., \& {Colless}, M. 2001, \apjl, 548, L139

\bibitem[{{Gerhard} {et~al.}(2001){Gerhard}, {Kronawitter}, {Saglia}, \&
  {Bender}}]{Gerhard:2001}
{Gerhard}, O., {Kronawitter}, A., {Saglia}, R.~P., \& {Bender}, R. 2001, \aj,
  121, 1936

\bibitem[{{Goudfrooij} {et~al.}(2001){Goudfrooij}, {Mack}, {Kissler-Patig},
  {Meylan}, \& {Minniti}}]{Goudfrooij:2001}
{Goudfrooij}, P., {Mack}, J., {Kissler-Patig}, M., {Meylan}, G., \& {Minniti},
  D. 2001, \mnras, 322, 643

\bibitem[{{Hui} {et~al.}(1995){Hui}, {Ford}, {Freeman}, \& {Dopita}}]{Hui:1995}
{Hui}, X., {Ford}, H.~C., {Freeman}, K.~C., \& {Dopita}, M.~A. 1995, \apj, 449,
  592

\bibitem[{{Kormendy} \& {Tremaine}(2011)}]{Kormendy:2011}
{Kormendy}, J. \& {Tremaine}, S. 2011, \araa, in prep.

\bibitem[{{Longhetti} {et~al.}(1998){Longhetti}, {Rampazzo}, {Bressan}, \&
  {Chiosi}}]{Longhetti:1998}
{Longhetti}, M., {Rampazzo}, R., {Bressan}, A., \& {Chiosi}, C. 1998, \aaps,
  130, 267

\bibitem[{{McNeil} {et~al.}(2010){McNeil}, {Arnaboldi}, {Freeman}, {Gerhard},
  {Coccato}, \& {Das}}]{McNeil:2010}
{McNeil}, E., {Arnaboldi}, M., {Freeman}, K., {et~al.} 2010, \aap, 518, A44+

\bibitem[{{M{\'e}ndez} {et~al.}(2001){M{\'e}ndez}, {Riffeser}, {Kudritzki},
  {Matthias}, {Freeman}, {Arnaboldi}, {Capaccioli}, \& {Gerhard}}]{Mendez:2001}
{M{\'e}ndez}, R.~H., {Riffeser}, A., {Kudritzki}, R.-P., {et~al.} 2001, \apj,
  563, 135

\bibitem[{{M{\'e}ndez} {et~al.}(2008){M{\'e}ndez}, {Teodorescu}, \&
  {Kudritzki}}]{Mendez:2008}
{M{\'e}ndez}, R.~H., {Teodorescu}, A.~M., \& {Kudritzki}, R.-P. 2008, \apjs,
  175, 522

\bibitem[{{Monet} {et~al.}(2003){Monet}, {Levine}, {Canzian}, {Ables}, {Bird},
  {Dahn}, {Guetter}, {Harris}, {Henden}, {Leggett}, {Levison}, {Luginbuhl},
  {Martini}, {Monet}, {Munn}, {Pier}, {Rhodes}, {Riepe}, {Sell}, {Stone},
  {Vrba}, {Walker}, {Westerhout}, {Brucato}, {Reid}, {Schoening}, {Hartley},
  {Read}, \& {Tritton}}]{Monet:2003}
{Monet}, D.~G., {Levine}, S.~E., {Canzian}, B., {et~al.} 2003, \aj, 125, 984

\bibitem[{{Peng} {et~al.}(2004){Peng}, {Ford}, \& {Freeman}}]{Peng:2004}
{Peng}, E.~W., {Ford}, H.~C., \& {Freeman}, K.~C. 2004, \apj, 602, 685

\bibitem[{{Proctor} {et~al.}(2009){Proctor}, {Forbes}, {Romanowsky}, {Brodie},
  {Strader}, {Spolaor}, {Mendel}, \& {Spitler}}]{Proctor:2009}
{Proctor}, R.~N., {Forbes}, D.~A., {Romanowsky}, A.~J., {et~al.} 2009, \mnras,
  398, 91

\bibitem[{{Quinn} \& {Zurek}(1988)}]{Quinn:1988}
{Quinn}, P.~J. \& {Zurek}, W.~H. 1988, \apj, 331, 1

\bibitem[{{Sch{\"o}nberner} {et~al.}(2010){Sch{\"o}nberner}, {Jacob}, {Sandin},
  \& {Steffen}}]{Schonberner:2010}
{Sch{\"o}nberner}, D., {Jacob}, R., {Sandin}, C., \& {Steffen}, M. 2010, \aap,
  523, A86+

\bibitem[{{Schuberth} {et~al.}(2010){Schuberth}, {Richtler}, {Hilker},
  {Dirsch}, {Bassino}, {Romanowsky}, \& {Infante}}]{Schuberth:2010}
{Schuberth}, Y., {Richtler}, T., {Hilker}, M., {et~al.} 2010, \aap, 513, A52+

\bibitem[{{Schwarz}(1978)}]{Schwarz:1978}
{Schwarz}, G. 1978, The Annals of Statistics, 6, 461

\bibitem[{{Schweizer}(1980)}]{Schweizer:1980}
{Schweizer}, F. 1980, \apj, 237, 303

\bibitem[{{Skrutskie} {et~al.}(2006){Skrutskie}, {Cutri}, {Stiening},
  {Weinberg}, {Schneider}, {Carpenter}, {Beichman}, {Capps}, {Chester},
  {Elias}, {Huchra}, {Liebert}, {Lonsdale}, {Monet}, {Price}, {Seitzer},
  {Jarrett}, {Kirkpatrick}, {Gizis}, {Howard}, {Evans}, {Fowler}, {Fullmer},
  {Hurt}, {Light}, {Kopan}, {Marsh}, {McCallon}, {Tam}, {Van Dyk}, \&
  {Wheelock}}]{Skrutskie:2006}
{Skrutskie}, M.~F., {Cutri}, R.~M., {Stiening}, R., {et~al.} 2006, \aj, 131,
  1163

\bibitem[{{Spitler} {et~al.}(2006){Spitler}, {Larsen}, {Strader}, {Brodie},
  {Forbes}, \& {Beasley}}]{Spitler:2006}
{Spitler}, L.~R., {Larsen}, S.~S., {Strader}, J., {et~al.} 2006, \aj, 132, 1593

\bibitem[{{Thomas} {et~al.}(2007){Thomas}, {Saglia}, {Bender}, {Thomas},
  {Gebhardt}, {Magorrian}, {Corsini}, \& {Wegner}}]{Thomas:2007}
{Thomas}, J., {Saglia}, R.~P., {Bender}, R., {et~al.} 2007, \mnras, 382, 657

\bibitem[{{Tonry} {et~al.}(2001){Tonry}, {Dressler}, {Blakeslee}, {Ajhar},
  {Fletcher}, {Luppino}, {Metzger}, \& {Moore}}]{Tonry:2001}
{Tonry}, J.~L., {Dressler}, A., {Blakeslee}, J.~P., {et~al.} 2001, \apj, 546,
  681

\bibitem[{{Treu} \& {Koopmans}(2004)}]{Treu:2004}
{Treu}, T. \& {Koopmans}, L.~V.~E. 2004, \apj, 611, 739

\bibitem[{{van der Kruit} \& {Freeman}(2011)}]{van-der-Kruit:2011}
{van der Kruit}, P.~C. \& {Freeman}, K.~C. 2011, \araa, 49, 301

\bibitem[{{Weijmans} {et~al.}(2009){Weijmans}, {Cappellari}, {Bacon}, {de
  Zeeuw}, {Emsellem}, {Falc{\'o}n-Barroso}, {Kuntschner}, {McDermid}, {van den
  Bosch}, \& {van de Ven}}]{Weijmans:2009}
{Weijmans}, A.-M., {Cappellari}, M., {Bacon}, R., {et~al.} 2009, \mnras, 1006

\bibitem[{{Woodley} {et~al.}(2010){Woodley}, {G{\'o}mez}, {Harris}, {Geisler},
  \& {Harris}}]{Woodley:2010}
{Woodley}, K.~A., {G{\'o}mez}, M., {Harris}, W.~E., {Geisler}, D., \& {Harris},
  G.~L.~H. 2010, \aj, 139, 1871

\bibitem[{{Wozniak} {et~al.}(1995){Wozniak}, {Friedli}, {Martinet}, {Martin},
  \& {Bratschi}}]{Wozniak:1995}
{Wozniak}, H., {Friedli}, D., {Martinet}, L., {Martin}, P., \& {Bratschi}, P.
  1995, \aaps, 111, 115

\end{thebibliography}
\bibliographystyle{aa}
\end{document}